\documentclass[preprint2,aastex]{emulateapj}
\usepackage{apjfonts}

\usepackage{graphics}

\newcommand{\etal}{et al.}

\newcommand{\app}{Astropart. Phys.}
\newcommand{\NIMA}{Nucl. Instr. Meth. A}

\newcommand{\degrees}{\ensuremath{^\circ}}
\newcommand{\tgg}{\ensuremath{\tau_{\gamma \gamma}}}

\begin{document}


\title{High-Energy Gamma-Ray Observations of W Comae with STACEE}

\author{\mbox{R. A. Scalzo}\altaffilmark{1,9},
        \mbox{L. M. Boone}\altaffilmark{2,10},
        \mbox{D. Bramel}\altaffilmark{3},
        \mbox{J. Carson}\altaffilmark{4},
        \mbox{C. E. Covault}\altaffilmark{5},
        \mbox{P. Fortin}\altaffilmark{6},
        \mbox{G. Gauthier}\altaffilmark{6},
        \mbox{D. M. Gingrich}\altaffilmark{7,8},
        \mbox{D. Hanna}\altaffilmark{6},
        \mbox{A. Jarvis}\altaffilmark{4},
        \mbox{J. Kildea}\altaffilmark{6},
        \mbox{T. Lindner}\altaffilmark{6},
        \mbox{C. Mueller}\altaffilmark{6},
        \mbox{R. Mukherjee}\altaffilmark{3},
        \mbox{R. A. Ong}\altaffilmark{4},
        \mbox{K. J. Ragan}\altaffilmark{6},
        \mbox{D. A. Williams}\altaffilmark{2},
        and \mbox{J. Zweerink}\altaffilmark{4} \\
        (The STACEE Collaboration)}
\altaffiltext{1}{Enrico Fermi Institute,
                 University of Chicago,
                 5640 S. Ellis Ave., Chicago, IL 60637}
\altaffiltext{2}{Santa Cruz Institute for Particle Physics,
                 University of California at Santa Cruz,
                 1156 High Street, Santa Cruz, CA 95064}
\altaffiltext{3}{Department of Physics and Astronomy,
                 Barnard College and Columbia University,
                 New York, NY 10027}
\altaffiltext{4}{Department of Physics and Astronomy,
                 University of California at Los Angeles,
                 Box 951547, Knudsen Hall, Los Angeles, CA 90095-1547}
\altaffiltext{5}{Department of Physics,
                 Case Western Reserve University,
                 10900 Euclid Ave., Cleveland, OH 44106}
\altaffiltext{6}{Department of Physics,
                 McGill University,
                 3600 University Street, Montreal, QC H3A 2T8, Canada}
\altaffiltext{7}{Centre for Subatomic Research, University of Alberta,
                 Edmonton, AB T6G 2N5, Canada}
\altaffiltext{8}{Tri-Universities Meson Facility,
                 Vancouver, BC V6T 2A3, Canada}
\altaffiltext{9}{present address:  Lawrence Berkeley National Laboratory,
                 MS 50R5008, 1 Cyclotron Road, Berkeley, CA 94720}
\altaffiltext{10}{present address:  Department of Physics,
                  The College of Wooster, Wooster, OH 44691}

\begin{abstract}
We report on observations of the blazar W Comae (ON+231) with the
Solar Tower Atmospheric Cherenkov Effect Experiment (STACEE),
a wavefront-sampling atmospheric Cherenkov telescope, in the spring of 2003.
In a data set comprising 10.5 hours of ON-source observing time,
we detect no significant emission from W Comae.
We discuss the implications of our results in the context of the composition
of the relativistic jet in W Comae, examining both leptonic and hadronic
models for the jet.
We derive 95\% confidence level upper limits on the flux at the level of
1.5--3.5$\times 10^{-10}$ cm$^{-2}$ s$^{-1}$ above 100 GeV
for the leptonic models,
or 0.5--1.1$\times 10^{-10}$ cm$^{-2}$ s$^{-1}$ above 150 GeV
for the hadronic models.
\end{abstract}



\section{Introduction}

The current catalog of extragalactic TeV gamma-ray sources consists 
of blazars, which are among the brightest and most rapidly variable objects
in the sky.  The unusual observational properties of blazars are usually
explained in terms of relativistic bulk motion of the emitting region
in a jet parallel to the line of sight \citep{up95}.  The continuum emission
of blazars is typically nonthermal and contains two broad peaks, one at lower
energies (radio to X-ray) and one at higher energies (keV to TeV).
Although the low-energy peak is generally assumed to result from synchrotron
radiation from high-energy electrons in the jet, the mechanism for the
production of the high-energy radiation is still a subject of debate,
and several competing models exist.

In ``leptonic'' jet models, the high-energy radiation is produced by inverse
Compton scattering of low-energy photons from a population of electrons
and/or positrons;
these models are favored for the well-studied blazars Mrk 421 and 501
\citep{ca99,konopelko03}.  In the synchrotron self-Compton model (SSC),
a single population of electrons and positrons produce the target photon
field via synchrotron radiation, and subsequently upscatter the synchrotron
photons to gamma-ray energies \citep{bm96}.  There may also
be an external Compton (EC) component of target photons from the accretion
disk or ambient medium \citep{dsm92,sbr94}.  Alternatively, in ``hadronic''
jet models, protons play a central role.  In hadronic models the high-energy
radiation is attributed to photomeson processes \citep{mannheim93}
or synchrotron radiation from protons or muons, as in the
synchrotron proton blazar (SPB) model \citep{mp00,aharonian00,mucke03}.
The study of hadronic models was originally motivated by the hypothesis
that the highest-energy cosmic rays are produced in blazar jets
\citep{mannheim95}.

The BL Lac object W Comae (ON+231) may provide an excellent test case for
hadronic jet models \citep{bmr02}.  The transition between the low-energy and
high-energy peaks in the continuum of W Comae appears clearly in X-ray data
taken by the BeppoSAX satellite \citep{tagliaferri00}.
These high-quality observations of the transition region place
tight constraints on leptonic models, requiring the predicted
gamma-ray emission to cut off sharply above 100 GeV.
In contrast, hadronic models may allow for significant emission above 100 GeV.
Observations by the EGRET detector aboard the
Compton Gamma Ray Observatory show a hard power law spectrum
(photon spectral index $\alpha = 1.73 \pm 0.18$) extending up to about
10 GeV with no sign of any cutoff \citep{hartman99}.
Yet the object has not been detected at energies above 300 GeV,
despite repeated observation by the Whipple 10-m instrument \citep{horan03}.
At a redshift of $0.1$, absorption of gamma rays in this energy range
by pair production \( \gamma \gamma \longrightarrow e^+e^- \)
against the extragalactic background light (EBL) may be significant, but only
at energies above about 500 GeV \citep{primack99,primack01,ms01,aharonian01},
so the intrinsic emission spectrum of W Comae should be directly observable
at energies lower than this.

The Solar Tower Atmospheric Cherenkov Effect Experiment (STACEE) currently
operates above an energy threshold of about 100 GeV for gamma rays.
W Comae has been observed in the past with previous versions of the STACEE
detector \citep{theoret-thesis}.
New observations of W Comae were carried out in the spring of 2003 by the
most recent version of the detector, STACEE-64.


\section{The STACEE detector}

STACEE is an atmospheric Cherenkov telescope which uses as its primary optic
an array of solar mirrors (heliostats) at the National Solar
Thermal Test Facility (NSTTF) in Albuquerque, NM.  STACEE uses the
independently steerable heliostats to collect Cherenkov light from extensive
air showers initiated by astrophysical gamma rays.  Secondary mirrors are
used to image the heliostat field onto a camera of photomultiplier tubes
(PMTs), producing a one-to-one mapping between heliostats and PMTs.
Optical concentrators based on the DTIRC design \citep{nwog87} widen
the aperture of each PMT from 5 cm to 11 cm and restrict its field of view
to reduce the number of night sky background (NSB) photons detected.
High-speed electronics measure the charge and relative arrival times of the
PMT pulses, and impose a coincidence trigger.  The STACEE detector is thus
a \emph{wavefront-sampling} detector, measuring the intensity and arrival
time of the narrow wavefront of Cherenkov light at different spots on the
ground.  These measurements enable the offline reconstruction of the primary
energy and shower arrival direction.

STACEE achieves a low energy threshold ($\sim 100$ GeV)
for the detection of gamma rays (compared with contemporary single-dish
imaging Cherenkov telescopes) primarily because of the extremely large
available mirror area ($A \sim 10^3$ m$^2$);
the energy threshold scales approximately as $A^{-1/2}$ \citep{weekes88}.
Other instruments using a similar solar tower design include the
Cherenkov Low Energy Sampling and Timing Experiment (CELESTE)
in Themis, France \citep{denaurois02},
and the Solar Two detector in Barstow, CA \citep{tripathi02}.
These are the only experiments currently operating in the Northern Hemisphere
which have their peak sensitivity to gamma rays below 300 GeV.

STACEE has undergone a series of upgrades and improvements over its lifetime.
The prototype experiment, called STACEE-32, used 32 heliostats with a total
mirror area of 1200 m$^2$ \citep{stacee32nim}, and successfully detected
gamma-ray emission from the Crab Nebula at a peak energy of 190 GeV
\citep{osercrab}.  An upgraded detector called STACEE-48 used an additional
16 heliostats (for a total of 48) and employed improved high-speed
programmable delay and trigger electronics \citep{maddog}; it detected
gamma-ray emission from Mrk 421 during a period of intense flaring activity
\citep{boonemrk421}, as part of a multi-wavelength variability study.
The final stage of construction,
called STACEE-64, was completed in the summer of 2001.  An additional 16
heliostats were instrumented, for a total of 64 heliostats with over 2300
m$^2$ of mirror area, filling in the geographical front and center of the
heliostat field.  The charge-integrating ADCs used by prior versions of
STACEE were also removed, and replaced with waveform digitizers or
``flash ADCs'' (FADCs).  A brief overview of these improvements follows;
additional detail is available elsewhere \citep{mythesis}.

\subsection{Optics upgrades}

\begin{figure}
\plotone{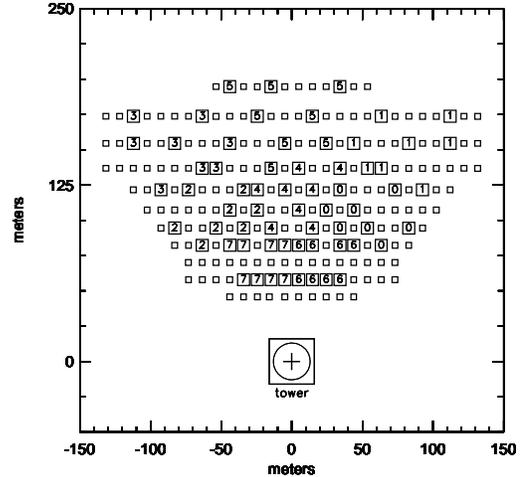}
\caption[Current state of the STACEE heliostat field]
{Map of the heliostat field used by the STACEE experiment, showing updates
to the experiment over time.  Heliostats are numbered according to the
trigger subcluster to which they belong.  Clusters 0, 1, 2, and 3 correspond
to the ``east'' and ``west'' cameras; most of these heliostats were used
in the STACEE-32 prototype.  Clusters 4 and 5 correspond to the ``north''
camera which was added for STACEE-48.  Clusters 6 and 7 correspond to the
new STACEE-64 ``southeast'' and ``southwest'' cameras.  The base of the
NSTTF solar tower, which houses the STACEE optics and electronics,
marks the origin of the coordinate system.}
\label{fig:birdsI-st64apj}
\end{figure}

Figure \ref{fig:birdsI-st64apj} shows the NSTTF heliostat field, with the
STACEE heliostats indicated.  The heliostats are chosen to provide the most
uniform ground coverage possible, subject to constraints imposed by the
crowding of heliostat images (and PMT apertures) in the focal plane of each
secondary mirror.  In addition to the three secondaries (1.9 m diameter)
used by STACEE-48, two additional secondaries (1.1 m diameter) were built
for STACEE-64 to allow use of additional heliostats at the front of the
field.  A set of 16 additional camera elements (PMT + DTIRC + housing)
were instrumented
for the new heliostats.  All 64 camera elements were carefully calibrated
in the lab and redistributed among the cameras according to measured
differences in quantum efficiencies, in order to equalize the optical
throughput on all detector channels.

Improved understanding of the heliostat optics via Monte Carlo simulations
and measurements of the heliostat point-spread function using a CCD camera
prompted adjustments to the optical elements.  Each STACEE heliostat
has 25 mirror facets which can be independently focused and aligned
\citep{stacee32nim}; the point spread function is periodically evaluated
and adjustments made, if necessary, using a laser look-back method.
Prior to the W Comae
observing campaign, the facets on many heliostats were re-adjusted to fix
the overall optical axis within the heliostat body frame to a standard
position, thereby improving pointing accuracy.  The absolute pointing of each
heliostat was calibrated to an accuracy of 0.05\degrees\ using drift scans
of bright stars.

\subsection{Trigger}

STACEE uses a two-level coincidence trigger described in detail in
\citet{maddog}.  PMT camera elements are grouped into subclusters of
eight PMTs each.  The signal from each PMT is AC-coupled, amplified,
and discriminated before entering the subcluster trigger electronics.
The coincidence trigger accounts for channel-to-channel delays among the
discriminated PMT signals, which arise from the geometry of the shower
wavefront and the differences in propagation times from different parts
of the heliostat field; the programmed delays are accurate to 1 ns.
A coincidence test among delayed PMT signals is then made at the
subcluster level, and a further coincidence among subclusters is required
to trigger event readout.
The number of coincident PMTs in a subcluster, and the number of
coincident subclusters, are chosen to optimize the quality factor for
the rejection of hadronic air showers according to Monte Carlo simulations.
The discriminator threshold is then set at a level which
makes the overall event trigger rate from chance coincidences of pulses due
to fluctuations of NSB photons negligible
(less than 0.2 min$^{-1}$).

The STACEE-64 trigger topology is shown in Figure \ref{fig:birdsI-st64apj}.
For the W Comae observations,
the discriminator threshold was set to 140 mV (about 5.5 photoelectrons);
five out of eight PMTs in a subcluster were required to cross threshold
within a narrow coincidence window to generate a subcluster trigger.
The precise width of this coincidence window as applied to a given series of
pulses varies between 8 and 24 ns due to the implementation of the trigger
logic \citep[see][]{maddog}; the mean width for a two-channel coincidence
is 16 ns.  Five out of eight subcluster triggers within a window of 16 ns
were necessary to trigger event readout.

\subsection{FADCs}


The FADCs represent a major upgrade to the STACEE experiment.  Access to
the full digitized PMT waveform allows not only more accurate measurements
of the timing and intensity of the wavefront, but also measurement
of charge-timing correlations, such as the distribution of Cherenkov photon
arrival times at each heliostat.
Various new methods which use FADC data to reject hadronic events
are currently under study by the STACEE collaboration
\citep{scalzo03,zweerink03}.
The FADCs are also routinely used to calibrate and monitor the gains of the
STACEE PMTs using a custom-designed laser calibration system
\citep{stlasernim}.

\begin{figure}
\plotone{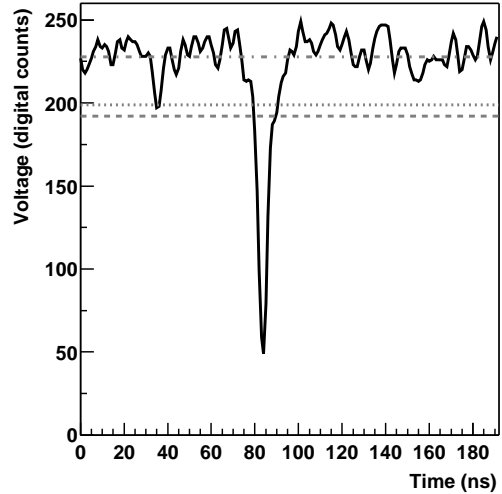}
\caption[Digitized PMT signal]
{Digitized PMT signal from a typical air shower event.
Dot-dash line:  FADC baseline.
Dashed line:  Discriminator threshold.
Dotted line:  3 standard deviations above baseline for NSB fluctuations.}
\label{fig:fadctrace}
\end{figure}

The FADCs used by STACEE-64 are a commercial system produced by Acqiris, Inc.
The system comes in modular pieces, with four channels per board.  Up to
six boards (24 channels) fit into a special crate, which runs the Linux
operating system.  A real-time Linux driver for the system was developed by
our group.  Each FADC channel samples at 1 GS/s with a dynamic range of
1 V during normal astronomical observations.  A sampled PMT
signal from an actual Cherenkov event is shown in Figure \ref{fig:fadctrace}.
The zero points of the FADC inputs are calibrated to a precision of 1 mV RMS,
and the channel-to-channel gains of the system are equalized to within 0.5\%.

In the 2001--2002 observing season, 32 FADC channels
were instrumented, each sampling the signal from two PMT channels chosen so
that the incoming Cherenkov signals would be well-separated in time.
By the 2002-2003 observing season, 64 FADC channels were instrumented so that
each FADC sampled the signal from exactly one PMT.


\section{Observations and Quality Cuts}


STACEE-64 was used to observe W Comae in the spring of 2003, from late March
to early June.  To maximize sensitivity and robustness to systematic effects,
STACEE operates in an ``ON-OFF'' integration mode \citep{osercrab}.
In ON-OFF observation, each gamma-ray source is assigned another
region of the sky at the same declination which contains no known gamma-ray
source.  The two regions are usually separated by 30 minutes in right
ascension.  STACEE then tracks the OFF region immediately before or
after the ON region, so that the same range of azimuth and elevation are
observed, and the detector sensitivity
(which depends on azimuth and elevation)
is matched in both halves of the pair.
Thirty-four ON-OFF pairs were taken on W Comae,
for a total of 13.5 hours of ON-source observing time.

Several quality cuts were then imposed on the data set, as described below.
These quality cuts are to be understood as ``pairwise time cuts'',
in the sense that if data in a certain time interval during one run are
flagged and removed from the data set, then data in the corresponding time
interval in the other run of the pair
(i.e., spanning the same local horizon coordinates) are also removed.

The control system for the NSTTF heliostats continually records status
information which can be used to pinpoint malfunctions on particular
heliostats.  This status information is logged nightly and merged with the
regular STACEE data product.  Portions of runs during which any heliostat
malfunctioned were flagged and removed from the data set.

As observed in \citet{boonemrk421}, the subcluster trigger rates are very
sensitive to fluctuations in sky conditions, since they vary as a high power
of the discriminator rates on PMTs within the cluster, which in turn are
dominated by fluctuations of NSB photons.
Large deviations of the subcluster trigger rates from a smooth linear trend
have been seen to correlate strongly with unstable weather conditions.
After heliostat cuts, an iterative procedure was used to identify and flag
affected portions of the data.  Each pair was divided into 30-s time bins,
and a least-squares fit to the time-averaged subcluster rates performed
repeatedly, excluding outliers beyond a certain tolerance at each iteration.
In order for a bin to merit exclusion from the data set, deviations in the
rates had to appear for at least four out of eight subclusters.
After undergoing this process, if a pair of data had less than a minute
of data remaining, this pair was simply excluded from further analysis.

Additionally, some data were removed because of malfunctions
in the data acquisition program resulting in the loss or corruption of
FADC data, which is needed for the final analysis (see below).
About 2\% of the original data set was removed for this reason.

Finally, even though the pairwise time cuts equalize the observing times
in the ON-source and OFF-source data sets, a livetime correction is
necessary.  The time required for the data acquisition software to read out
each triggered event creates dead time within the observing intervals,
which is rate-dependent and measured electronically by the detector
as the data are acquired.

After all quality cuts, a total of 10.5 hours of ON-source observing time
remained, distributed among 32 ON-OFF pairs (see Table \ref{tbl:padding}).
Using the \citet{lima} expression for the significance gives a (raw)
positive significance of 4.6 standard deviations,
at a livetime-corrected excess event rate of 4.8 $\pm$ 1.1 min$^{-1}$.


\section{Sky Brightness Corrections}

Before deriving an integral flux or upper limit, it is very important to
correct the raw trigger rate for any differences in NSB fluctuations between
the ON-source and OFF-source fields.
NSB photons entering the detector frequently
promote sub-threshold hadronic air showers above the detection threshold,
even under the best conditions.  The promotion rate increases as the
sky brightness increases, so that if the ON-source field is slightly brighter
than the OFF-source field, either due to stars or to subtle variations in
atmospheric conditions, the resulting excess in the \emph{hadronic}
trigger rate can mimic a gamma-ray excess.  Similarly, if the OFF-source
field is brighter than the ON-source field, an existing gamma-ray excess may
be masked; a net deficit may even appear.  An accurate measurement of the
true gamma-ray rate therefore requires a correction for the promotion effect.

We have developed three independent methods to correct for the effects
of varying levels of NSB light on the coincidence trigger.
We find that the excess we see from W Comae is consistent with the promotion
of events representing only hadronic showers, with no additional gamma-ray
component.

\subsection{Direct measurement of the hadronic promotion trend}

One method of correcting for field brightness effects was demonstrated by
the analysis of the STACEE-48 observations of Mrk 421 \citep{boonemrk421}.
The excess brightness in the STACEE ON-source field for Mrk 421 was due
mainly to a single bright star, HD 95934 (magnitude 6.16 in the $B$-band).
No FADC data were available at that time, but the promotion rate was
estimated simply by taking ON-OFF pairs on another star, HIP 80460,
with a similar $B$-band magnitude.  The measured promotion rate
was then subtracted from the total excess to give the gamma-ray excess.

\begin{figure}
\plotone{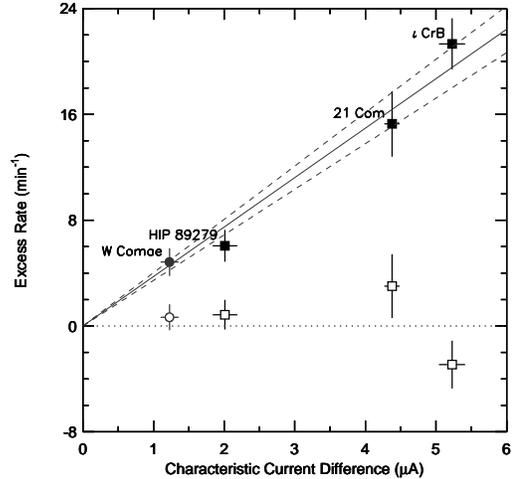}
\caption[Promotion rate vs. current]
{Excess event trigger rates as a function of the characteristic current
difference for sources observed in 2002--2003 by STACEE-64.  Squares:
star observations; circle:  W Comae.  Filled symbols show the raw excess,
and open symbols show the excess after the dynamic threshold brightness
correction.  The lines show the best-fit linear
trend constrained to go through the origin (solid) with 68\% confidence errors
on the slope (dashed), and the zero-rate baseline (dotted).}
\label{fig:RvI}
\end{figure}

Similar observations have been made for STACEE-64 and are depicted in
Figure \ref{fig:RvI}.  Three stars of different magnitudes
and suitable declinations (21 Com, $\iota$ CrB, and HIP 89279) were
selected and observed as if they were gamma-ray sources.  The excess rate
is plotted as a function of a weighted average current difference,
\begin{equation}
\Delta I = \frac{\sum_{j=1}^{64} \left< w_j \, \Delta I_j \right>}
   {\sum_{j=1}^{64} \left< w_j \right>},
\end{equation}
where $\Delta I_j$ is the ON-OFF difference in anode current on PMT $j$,
$w_j$ is the measured fraction of triggered events in which PMT $j$
crossed threshold, and the angle brackets denote time-averaging over
the entire data set.  The promotion trend is well fit
($\chi^2/\mathrm{d.o.f.} = 1.04$) by a straight line
with slope $3.7 \pm 0.3$ min$^{-1}$ $\mu$A$^{-1}$;
it represents a zero-signal baseline for gamma-ray sources observed by STACEE.

The raw rate measurement for W Comae is also shown on the plot,
and it is consistent with the
promotion trend.  Subtracting the expected number of promotion events from
the W Comae excess based on its characteristic current leaves a net excess of
$0.2 \pm 1.2$ min$^{-1}$.

The direct measurement of the promotion trend is intuitive and
straightforward, and can be used to quickly assess how much of an observed
excess is due to promotion.  However, the promotion trend as measured
reflects the observing season on average, and in particular observing
conditions during the star runs, rather than conditions which might pertain
for individual runs in the data set for a particular gamma-ray source.
An event-by-event, channel-by-channel treatment is therefore more
desirable when deriving scientific results such as fluxes or upper limits.
A sizable data set of star observations proved quite valuable for testing
such brightness correction techniques.

\subsection{Software padding using waveform libraries}


Another technique, commonly referred to as \emph{padding},
has been in use in some form by atmospheric Cherenkov telescopes since
their inception.  For example, in the early days of the Whipple 10-m
imaging telescope, NSB conditions were equalized in
hardware by using a small LED on the face of each PMT in a feedback loop
\citep{weekes89}.  Later the equalization was done in software, by adding
random Gaussian deviates to the recorded ADC values to mimic fluctuations
from increased NSB levels \citep{cawley93}, and by then
reimposing an additional selection criterion similar to the hardware trigger.
For STACEE, the trigger depends not only on charge distributions but also on
relative timing, since the light from an air shower reaches different parts
of the detector at different times.  Knowledge of the PMT waveforms
as provided by the FADCs is therefore necessary to do software padding.

A software padding scheme has been implemented and tested by STACEE
as follows:  FADC waveforms corresponding to varying NSB levels
were sampled from 16 of the PMTs \emph{in situ} under controlled conditions.
The waveforms were characterized by their RMS fluctuations and stored in
a library to be used in padding.  In the padding analysis, these library
waveforms were added to the waveforms in the observation data set so as to
equalize the RMS fluctuations in both runs of a pair, as measured by the
400-ns section of the waveform immediately preceding the event
trigger.  An offline reimposition of the trigger required five out of eight
PMTs in a subcluster, and five out of eight subclusters, to fire within a
coincidence window of width 16 ns.

\begin{figure}
\center
\plotone{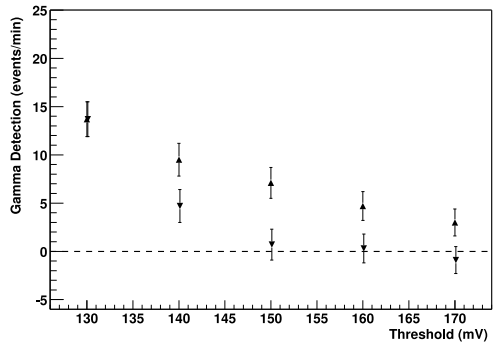}
\caption[Promotion excess vs. FADC analysis threshold for library padding]
{Excess event trigger rate as a function of FADC analysis threshold
in the library padding analysis of a subset of data taken on the star
$\iota$ CrB.  The trigger condition is described in the text.
Upright triangles:  No padding of traces before retriggering.
Inverted triangles:  Traces are padded to equalize RMS before retriggering.
Dashed line:  zero promotion excess.
The figure shows that the minimum threshold necessary to remove all
hadronic promotion events from the library-padded data set is 150 mV.}
\label{fig:tlthresh}
\end{figure}

When reimposing the trigger in software, the analysis threshold at which the
FADC waveforms are discriminated must be increased over the nominal hardware
discrimination threshold.  Adding background noise to waveforms from recorded
events in a low-noise run equalizes only the probability that an event will
pass additional offline cuts relative to an event in the corresponding
high-noise run.  It is impossible to recover events which \emph{would} have
appeared in the low-noise data set if the additional noise had been physically
present at the outset.  A subset of the ON-OFF data for $\iota$ CrB were used
to optimize the offline analysis threshold (see Figure \ref{fig:tlthresh}).
The analysis threshold was set to 150 mV for all FADC channels based on the
results of this optimization.

The results of the library padding brightness correction procedure
are shown in Table \ref{tbl:padding}.  For all three stars and for
W Comae, the ON-OFF difference drops from a significant excess in the
uncorrected data to an insignificant difference in the corrected data.
In particular, the significance drops from 4.6 to 0.9 standard deviations
for W Comae.

\subsection{Dynamic thresholds}



Although software padding has clearly been successful in eliminating the
promotion excess, it has the drawback that the software discriminator
threshold used to analyze FADC waveforms, and hence the energy threshold
for the analysis, must be raised significantly.  The desire to
maintain a low discriminator threshold in the offline analysis of FADC data
formed the motivation for investigating what we call the
\emph{dynamic thresholds} technique.

In the absence of NSB fluctuations, the distribution of pulse heights for
pulses within the trigger coincidence window on each channel would depend
only on the intensity of Cherenkov light from air showers.  The presence of
NSB fluctuations on a channel has two effects on the pulse height spectrum.
First, the linear superposition of fluctuating background traces upon the
underlying pulses from Cherenkov photons produces an additional statistical
uncertainty in the pulse height measurement for each event, by changing the
pedestal from which the pulse height is measured.  NSB fluctuations therefore
smear out the underlying air shower pulse height distribution.
Second, NSB fluctuations occasionally cross the discriminator threshold
within the coincidence window even when no signal pulse is present,
resulting in a second component to the pulse height distribution.

The dynamic thresholds technique models both of these effects quantitatively
to predict changes in the analysis thresholds.
When comparing the ON-source and OFF-source runs of a pair, the run with
larger fluctuations in the baseline trace will have a higher promotion rate,
and therefore a lower effective discriminator threshold.  To compensate for
promotion, for each channel the threshold in the noisier run is raised so
that the rate of \emph{background} pulses (NSB plus hadronic Cherenkov light)
within the coincidence window, and with pulse height above the analysis
threshold, is the same in both halves of the pair.  When a coincidence
condition among pulses is imposed offline, this prescription should remove
a number of events equal to the expected number of promoted hadronic
shower events, while retaining sensitivity to any possible gamma-ray excess.

For the OFF-source run, only hadronic air showers and NSB photons contribute
to the pulse height distribution, so that the rate of pulses above the
analysis threshold can be directly measured.  The pulse height distribution
for the ON-source run, however, may also contain Cherenkov pulses from
gamma-ray air showers.  Thus, the form of the ON-source pulse height spectrum
\emph{without} the gamma-ray contribution must be predicted from
the measured OFF-source pulse height spectrum and measurements of the
NSB fluctuations.

\newcommand{\sV}[1]{\ensuremath{s_\mathrm{#1}(V)}}
\newcommand{\RV}[1]{\ensuremath{R_\mathrm{#1}(V)}}

\begin{figure}
\plotone{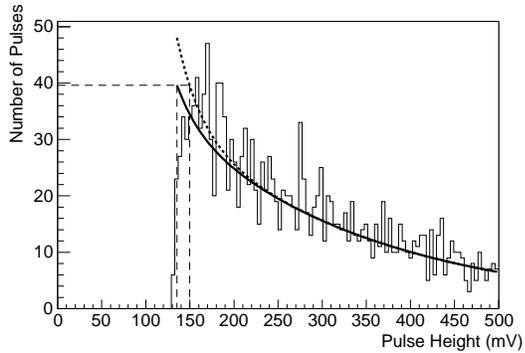}
\caption[Dynamic threshold calculation]
{Pulse height distribution of triggered events from a 10-minute segment of
a typical OFF-source run on the star $\iota$ CrB.  The best-fit form of
a Poisson distribution (expected NSB) plus an exponential is
shown (solid line), as well as the predicted form for the ON-source run
(dashed line) resulting from additional NSB fluctuations.
The ON-source prediction is normalized to the same detector livetime
as the OFF-source best fit.  Horizontal dashed line:  number of OFF-source
pulses per mV at the nominal hardware discriminator threshold.
Vertical dashed lines:  analysis thresholds ON and OFF necessary to
equalize the number of pulses above threshold over the course of the run.}
\label{fig:dynthresh-phc}
\end{figure}

We used a semi-analytic approach to predict the form of the ON-source
pulse height spectrum \citep[for details, see][]{mythesis}.
Histograms of individual waveform samples were accumulated using the part
of each waveform immediately preceding the event trigger.
These histograms characterized the distribution of fluctuations, denoted
\sV{ON}\ and \sV{OFF}, in the waveform baseline
(and hence in the effective discriminator threshold) due to NSB fluctuations
in each run.  The pulse height distribution for all pulses within the
coincidence trigger window for the OFF-source run,
denoted \RV{OFF}, was also measured.  We predicted the form of the ON-source
pulse height spectrum \RV{ON}\ as follows:
\begin{enumerate}
\item Fit \sV{OFF}, \sV{ON}\ and \RV{OFF} with smooth analytic forms
      to interpolate between thresholds.
\item Normalize \sV{OFF}\ with respect to \RV{OFF}\ using the measured
      OFF-source PMT rate (which is dominated by NSB fluctuations)
      and subtract it from \RV{OFF}\ to obtain a pulse
      height spectrum representing only pulses due to hadronic air showers.
\item Deconvolve this hadronic pulse height spectrum by \sV{OFF}\
      to obtain \RV{0}, the expected hadronic air shower pulse height
      spectrum in the absence of NSB fluctuations.
\item Convolve \RV{0}\ by \sV{ON}\ to account for the NSB fluctuations
      in the ON-source run.
\item Normalize \sV{ON}\ using the measured PMT rate ON-source and add
      it to the convolution of \RV{0}\ and \sV{ON}\ to obtain the
      predicted form of \RV{ON}.
\end{enumerate}
Figure \ref{fig:dynthresh-phc} shows an example of the results of this
procedure for one
STACEE channel in the data set for the star $\iota$ CrB.  A visual
inspection of Figure \ref{fig:dynthresh-phc} indicates that the threshold
on this channel must be increased by 15 mV.  This figure represents the
worst-case scenario, corresponding to conditions on one of the noisiest
channels during observations
of the brightest of the three padding stars.  For the W Comae observations,
the mean increase in the software discriminator threshold is approximately
2 mV, considerably smaller than the
10 mV adjustment on \emph{all} channels necessary for software padding.

The dynamic thresholds for a run were always calculated after quality cuts
to eliminate systematics due to rapidly fluctuating sky conditions.
To improve robustness against slow drifts in sky conditions, each 28-minute
run was partitioned into three equal segments 560 seconds in length.
The statistics within each segment were sufficient to establish the form of
the sampled distributions, except in cases where time cuts had already
removed most of the segment.  If statistics within a segment did not permit
a reliable prediction of the threshold adjustments, the segment and its
counterpart in the other half of the pair were removed from the data set.
Less than 3\% of the quality-cut data set was removed in this way;
the remaining data set comprised 10.3 hours of ON-source observing.

The results of the dynamic threshold analysis for W Comae, and for the three
stars observed by STACEE for promotion studies, are shown in
Table \ref{tbl:padding}.  As with library padding, no statistically
significant excess remains on any of the stars, or on W Comae, after the
dynamic threshold technique is applied to the raw data sets.


\begin{deluxetable}{lrrrrr}
\tablewidth{0pt}
\tablecolumns{6}
\tablecaption{Results of field brightness corrections \label{tbl:padding}}
\tablehead{
   \colhead{Source} &
   \colhead{$t_{ON}$\tablenotemark{a} (s)} &
   \colhead{$t_{OFF}$\tablenotemark{a} (s)} &
   \colhead{$N_{ON}$} & \colhead{$N_{OFF}$} &
   \colhead{Sig. ($\sigma$)\tablenotemark{b}}
}
\startdata
\sidehead{Quality cuts only:}
$\iota$ CrB & 11777.4 & 12064.9 &  75505 &  73060 & +10.99 \\
21 Com      &  6053.2 &  6148.7 &  32354 &  31299 &  +6.16 \\
HIP 89729   & 19087.2 & 19191.8 &  74823 &  73292 &  +5.03 \\
W Comae     & 37767.0 & 37955.5 & 219031 & 217085 &  +4.60 \\
\sidehead{Library padding:}
$\iota$ CrB & 11777.4 & 12064.9 &  52423 &  53367 &  +1.02 \\
21 Com      &  6053.2 &  6148.7 &  22859 &  23269 &  -0.23 \\
HIP 89729   & 19087.2 & 19191.8 &  53463 &  53678 &  +0.24 \\
W Comae     & 37767.0 & 37955.5 & 155722 & 156006 &  +0.88 \\
\sidehead{Dynamic thresholds:}
$\iota$ CrB & 11234.1 & 11508.5 &  58566 &  60556 &  -1.60 \\
21 Com      &  5522.8 &  5613.9 &  25172 &  25305 &  +1.25 \\
HIP 89729   & 18731.9 & 18835.3 &  61990 &  62064 &  +0.76 \\
W Comae     & 36982.9 & 37168.9 & 182433 & 182942 &  +0.67
\enddata
\tablecomments{Results of field brightness corrections for three stars
   of varying magnitude, and for W Comae.  The live times and raw excesses
   of events in the ON and OFF runs, and the corresponding significance,
   are shown for each technique.}
\tablenotetext{a}{Integrated live time after quality cuts and dead time
   correction.  Note that these represent equal amounts of actual observing
   time in the ON and OFF runs, but that differences in the trigger rate may
   lead to differences in the detector dead time between the ON and OFF runs.}
\tablenotetext{b}{Significance in standard deviations ($\sigma$).
   Calculated according to \citet{lima}.}
\end{deluxetable}


\section{Flux Limit Determination}



To calculate an upper limit on the flux of gamma rays from W Comae,
two main ingredients are necessary:  an assumption about the shape of the
spectrum, and knowledge of the detector response.  The latter was produced
from Monte Carlo simulations.  We use the CORSIKA simulation package
\citep{corsika} to model both hadronic and gamma-ray air showers.
The detector was modeled using software written by the collaboration,
including a full ray-trace of the detector optics and detailed simulations
of the discriminators, the delay and trigger systems, and the FADCs.
The full simulation chain
successfully reproduces important detector-related quantities, such as
the discriminator rate on each PMT, the trigger rate of cosmic-ray
air shower triggers at zenith, and the increase in the hadronic trigger rate
resulting from NSB fluctuations.

The response of the STACEE detector is characterized by the effective area,
defined as a function of energy by
\begin{equation}
A(E) = \int_G P(x,y,E) \, dx \, dy
\end{equation}
where $P(x,y,E)$ is the probability that a gamma-ray air shower with shower
core landing at position $(x,y)$ produces an event trigger, and $G$ is a
region on the ground (the $xy$-plane) containing all $(x,y)$ for which
$P(x,y,E) \neq 0$.  The functional form of the effective area
used to obtain the integral flux limits in this paper is an average of the
results of simulations done in different regions of the sky,
at a declination of +28.23\degrees\ spaced in 10\degrees\ increments in hour
angle, each weighted by its exposure in the 2003 data set.  The sensitivity
to very low-energy air showers is highest for a source at transit.
However, about 80\%
of the 2003 W Comae data were taken between hour angles of +20\degrees\ and
+40\degrees (zenith angles between 10\degrees\ and 26\degrees).
The large zenith angles result in a loss of sensitivity for low-energy air
showers with respect to that expected for the source at transit.
First, the Cherenkov photon density of the shower decreases, and the area
of the Cherenkov light pool increases, with increasing zenith angle.
Second, off-axis aberrations in the heliostat optics broaden the heliostat
point spread function for sky locations far from transit, decreasing
the overall optical throughput.

The effective area must also account for any energy dependence
introduced by the offline trigger cut used in the analysis.
Of the three analysis methods available to correct for field brightness
differences, the dynamic threshold technique maintains the best sensitivity
at low energies, and we use this technique to establish our upper limit.

In previous STACEE publications \citep{osercrab,boonemrk421},
the energy threshold has been quoted as the
energy at which the differential gamma-ray trigger rate
(effective area times expected photon flux from the assumed model)
reaches a maximum.  We continue to do so in this paper, quoting an energy
threshold for each assumed model accordingly.
We assume that the errors in the absolute throughput calibration of the
independent elements (heliostat optics, PMT quantum efficiency, etc.)
of the STACEE detector are Gaussian and uncorrelated.  These errors may then
be added in quadrature to obtain the systematic uncertainty in the absolute
energy scale calibration, which we estimate to be 20\%.
The systematic errors on the energy threshold are determined by rescaling
the energy scale at which the effective area is evaluated by $\pm 20\%$,
and calculating the resulting shift in the maximum of the differential
gamma-ray trigger rate.


\begin{figure}
\plotone{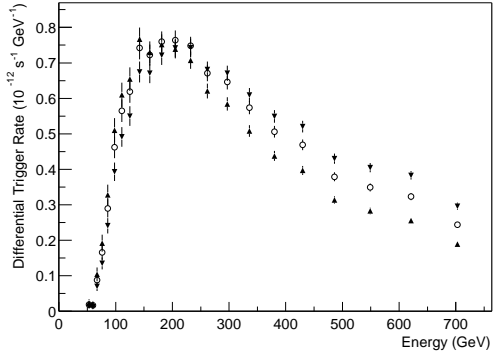}
\caption[STACEE-64 differential trigger rate]
{STACEE-64 simulated trigger rate vs. energy for a power law differential
flux spectrum of gamma rays with photon index
$\alpha = 1.55$ (inverted triangles),
$\alpha = 1.73$ (circles), and
$\alpha = 1.91$ (triangles).
All spectra are normalized to the same integral flux above 50 GeV.}
\label{fig:trigrate}
\end{figure}

Figure \ref{fig:trigrate} demonstrates the calculation of the energy
threshold for an extrapolation of the EGRET power law spectrum for W Comae
($\alpha = 1.73 \pm 0.18$), using the exposure-averaged effective area.
Curves representing the low and high values for the spectral index within
the EGRET experimental errors ($\alpha = 1.55$ or 1.91) are also shown,
normalized to the same integral flux above 50 GeV.
At transit, at a zenith angle of 6.73\degrees,
the energy threshold is about 120 GeV for $\alpha = 1.73$.
The energy threshold using the exposure-averaged effective area curve
is 170 GeV.  For the $\alpha = 1.73$ model, therefore, we quote our result
at an energy threshold of $170 \pm 40$ GeV.  This threshold changes
by about 20 GeV if the low or high value of $\alpha$ (1.55 or 1.91)
is used instead.

\begin{deluxetable}{lcrrr}
\tablewidth{0pt}
\tablecolumns{5}
\tablecaption{Integral flux upper limits for power law spectra
              \label{tbl:uplims-powlaws}}
\tablehead{ \colhead{Emission Model} &
            \colhead{\tgg\tablenotemark{a}} &
            \colhead{$E_\mathrm{thresh}$\tablenotemark{b}} &
            \colhead{95\% CL\tablenotemark{c}} &
            \colhead{$\Phi (> E_\mathrm{thresh})$\tablenotemark{d}} }
\startdata
 $\alpha = 1.55$ & low  & $191^{+44}_{-30}$ & 0.48 & 7.28 \\
 $\alpha = 1.55$ & high & $186^{+22}_{-29}$ & 0.52 & 3.87 \\
 $\alpha = 1.73$ & low  & $186^{+31}_{-30}$ & 0.49 & 2.23 \\
 $\alpha = 1.73$ & high & $170^{+26}_{-16}$ & 0.60 & 1.49 \\
 $\alpha = 1.91$ & low  & $168^{+38}_{-26}$ & 0.56 & 0.73 \\
 $\alpha = 1.91$ & high & $160^{+20}_{-23}$ & 0.65 & 0.52
\enddata
\tablecomments{Upper limits on the integral flux of gamma rays from W Comae,
   assuming that the differential flux of photons follows a power law
   ($\Phi(E) \propto E^{-\alpha}$).  Power laws were extrapolated from the
   Third EGRET Catalog \citep{hartman99} and corrected for EBL absorption.
   The mean spectral index $\alpha = 1.73$ is used, as well as low and high
   values consistent with the errors on the power-law fit to the EGRET data.} 
\tablenotetext{a}{Optical depth to pair production
   $\gamma \gamma \longrightarrow e^+e^-$ against the EBL from a redshift
   of $z = 0.102$, from \cite{aharonian01}.  The extreme low and high
   predictions for \tgg were used in this table.}
\tablenotetext{b}{STACEE energy threshold in GeV for the assumed model.
   The energy threshold is defined as the energy at which the differential
   trigger rate (effective area times gamma-ray flux) reaches a maximum.
   Systematic errors as shown represent propagation of the error in the
   calibration of the absolute energy scale using the mean effective
   area curve.}
\tablenotetext{c}{STACEE 95\% confidence level upper limit on the integral
   flux above the energy threshold,
   in units of 10$^{-10}$ cm$^{-2}$ s$^{-1}$.}
\tablenotetext{d}{Expected flux above the energy threshold according to
   the emission model, in units of 10$^{-10}$ cm$^{-2}$ s$^{-1}$.}
\end{deluxetable}

\begin{deluxetable}{lrrr}
\tablewidth{0pt}
\tablecolumns{4}
\tablecaption{Integral flux upper limits for leptonic models
              \label{tbl:uplims-leptonic}}
\tablehead{ \colhead{Emission Model} &
            \colhead{$E_\mathrm{thresh}$} &
            \colhead{95\% CL} &
            \colhead{$\Phi (> E_\mathrm{thresh})$} }
\startdata
SSC fit  1 & $113^{+25}_{-19}$ & 1.35 & 0.0278 \\
SSC fit  2 & $112^{+26}_{-18}$ & 1.39 & 0.0603 \\
SSC fit  3 & $108^{+26}_{-18}$ & 1.55 & 0.0876 \\
SSC fit  4 & $ 97^{+24}_{-13}$ & 2.12 & 0.139  \\
SSC fit  5 & $ 86^{+15}_{- 8}$ & 3.49 & 0.17   \\
SSC fit  6 & $ 95^{+23}_{-12}$ & 2.36 & 0.0276 \\
SSC fit  7 & $100^{+25}_{-15}$ & 1.94 & 0.331  \\
SSC+EC fit  8 & $112^{+26}_{-18}$ & 1.37 & 0.0303 \\
SSC+EC fit  9 & $110^{+26}_{-18}$ & 1.41 & 0.0344 \\
SSC+EC fit 10 & $108^{+27}_{-18}$ & 1.44 & 0.0467
\enddata
\tablecomments{Upper limits on the integral flux of gamma rays from W Comae,
   according to various leptonic (SSC, SSC+EC) models of the relativistic
   jet described in \citet{bmr02}, denoted BMR02 above.  Due to the sharp
   cutoff in the expected gamma-ray emission spectra above $\sim 100$ GeV,
   EBL absorption should be negligible for the leptonic models.  The table
   column headings are defined in the notes to
   Table \ref{tbl:uplims-powlaws}.}
\end{deluxetable}

\begin{deluxetable}{lcrrr}
\tablewidth{0pt}
\tablecolumns{5}
\tablecaption{Integral flux upper limits for hadronic models
              \label{tbl:uplims-hadronic}}
\tablehead{ \colhead{Emission Model} &
            \colhead{\tgg\tablenotemark{a}} &
            \colhead{$E_\mathrm{thresh}$} &
            \colhead{95\% CL} &
            \colhead{$\Phi (> E_\mathrm{thresh})$} }
\startdata
SPB fit 1 & low  & $127^{+22}_{-19}$ & 0.93 & 0.87 \\
SPB fit 1 & high & $126^{+16}_{-18}$ & 1.05 & 0.66 \\
SPB fit 2 & low  & $150^{+14}_{-16}$ & 0.72 & 2.48 \\
SPB fit 2 & high & $146^{+14}_{-16}$ & 0.77 & 1.91 \\
SPB fit 3 & low  & $164^{+29}_{-28}$ & 0.58 & 0.40 \\
SPB fit 3 & high & $157^{+22}_{-26}$ & 0.67 & 0.26 \\
SPB fit 4 & low  & $163^{+30}_{-27}$ & 0.59 & 0.44 \\
SPB fit 4 & high & $156^{+22}_{-26}$ & 0.69 & 0.30 \\
SPB fit 5 & low  & $160^{+40}_{-28}$ & 0.60 & 0.46 \\
SPB fit 5 & high & $141^{+36}_{-23}$ & 0.82 & 0.36
\enddata
\tablecomments{Upper limits on the integral flux of gamma rays from W Comae,
   according to various hadronic (SPB) models of the relativistic jet
   described in \citet{bmr02}, denoted BMR02 above.  The table column headings
   are defined in the notes to Table \ref{tbl:uplims-powlaws}.}
\end{deluxetable}

Tables \ref{tbl:uplims-powlaws},
\ref{tbl:uplims-leptonic}, and \ref{tbl:uplims-hadronic}
show upper limits from STACEE observations for
selected models of the gamma ray emission from W Comae \citep{bmr02}.
In each case the optical depth for pair production against the EBL
has been taken into account.  The models are based on
the semi-analytic approaches of \citet{primack99} and \citet{primack01},
which use a top-down, hierarchical treatment of structure formation as tuned
to fit recent EBL measurements.  The qualifiers ``low'' and ``high''
refer to the extreme low \citep[LCDM-Salpeter;][]{primack99}
and high \citep[Kennicutt;][]{primack01} predictions
for the optical depth to pair production at $z = 0.102$ from specific
instances of these semi-analytic models, as detailed in \citet{aharonian01}
and \citet{bmr02}.

It is apparent that, even when EBL absorption is taken into account,
the STACEE results place strong constraints on the extrapolation of
the EGRET power law to higher energies.  A steeper value of the spectral
index, within the uncertainties, is favored.  As noted by
\citet{bmr02}, however, the EGRET data were not taken simultaneously with
the measurements at other wavelengths.  The possible power law spectrum
may therefore represent a transient flaring state.  Moreover, the EGRET
power law is fit to a co-addition of many low significance
($\sim 2$ standard deviations) detections from different epochs,
with large uncertainties on the flux and spectral index.  In fact,
the leptonic model fits shown used only the Beppo-SAX X-ray data,
which is of higher quality;
no pure SSC model can fit both the X-ray and EGRET data simultaneously.

\begin{figure}
\plotone{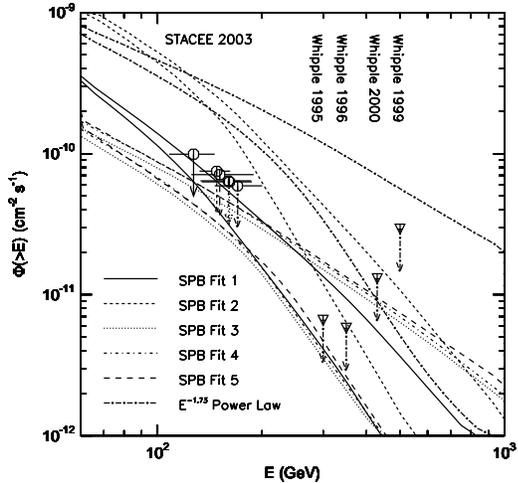}
\caption[STACEE integral flux limits on hadronic SPB models]
{Integral flux in cm$^{-2}$ s$^{-1}$ for each of the five SPB models
from \citet{bmr02}, and for the extrapolation of the best-fit EGRET
power law \citep{hartman99}.  Model predictions for both low and high
EBL optical depth are shown, along with STACEE upper limits for
mean EBL optical depth.  Upper limits from observations with the Whipple
10-meter from 1995--2000 are also shown \citep{horan03}.}
\label{fig:intfluxlimit}
\end{figure}

The STACEE observations are unable to place interesting constraints on most
of the leptonic models, with the expected fluxes being an order of magnitude
or more below the STACEE limit in each case.  The low predicted fluxes may
be anticipated from the sharp spectral cutoff in each model at or below
100 GeV, due to a similar sharp cutoff in the electron energy spectrum at
\( \gamma_{e,\mathrm{max}} \sim 5 \times 10^4 \).
Since the dominating factor is the availability of energy in relativistic
electrons, the cutoff energy in the modeled gamma-ray spectra is
approximately independent of the density of external target photons
in SSC+EC models.  The cutoff in the electron energy spectrum is
constrained tightly by the X-ray data, which lie between the falling edge
of the low-energy (electron synchrotron) peak in the spectrum and the
high-energy peak.

On the other hand, most of the hadronic models predict integral fluxes above
the energy threshold which are only slightly below the corresponding upper
limits from the STACEE observations.
In particular, we can exclude SPB model 2 at 95\% confidence.
Among all the hadronic models shown, model 2 has the
highest cutoff for the proton energy distribution
\( (\gamma_{p,\mathrm{max}} = 3 \times 10^9) \), yet the magnetic field,
the Doppler factor of the jet, etc., are comparable to the other models.

Figure \ref{fig:intfluxlimit} shows the STACEE upper limits for the hadronic
models in a graphical form.
The figure shows that STACEE is quite close to the sensitivity necessary
to begin to distinguish between the various hadronic models for W Comae.
However, the constraints provided by precise, simultaneous observations
at other wavelengths are necessary, in addition to STACEE observations,
because of the potential high-energy variability of blazars such as W Comae.
The constraints placed by the X-ray observations from SAX were instrumental
in shaping the predictions for the gamma-ray spectra for the hadronic models
presented here.  Figure \ref{fig:intfluxlimit} also shows
upper limits from high-energy observations made using the Whipple 10-meter
imaging Cherenkov telescope \citep{horan03} from 1995 to 2000.
(The energy thresholds for the Whipple limits were each calculated
assuming a power law with spectral index $\alpha = 2.5$, and do not yet
address specific emission models.)  Each instrument operates within a
different regime of sensitivity and energy, and the scientific payoff for a
variable target is maximized when observations are made simultaneously.
Further STACEE observations in the future, coupled with new offline
hadronic rejection techniques to improve the flux sensitivity, should allow
STACEE to further explore the parameter space for the hadronic models,
especially when contributing to a multi-wavelength monitoring campaign.



\section{Conclusions}

The STACEE-64 observations in the spring of 2003 were made at a lower
energy threshold than any other atmospheric Cherenkov telescope has
yet attained for W Comae.  STACEE detects no significant
emission from W Comae, resulting in 95\% CL upper limits on the integral flux
above this threshold in various hadronic emission models at the level of
$10^{-10}$ cm$^{-2}$ s$^{-1}$.
While leptonic models predict a flux which falls below this level,
extrapolations of the best-fit EGRET power law, and some synchrotron-proton
hadronic models, predict an integral gamma-ray flux above the
energy threshold close to, or exceeding, the upper limit from
STACEE observations.
Additional STACEE observations planned for the spring of 2004
ought either to exclude these models at a significantly higher confidence
level, or to detect gamma-ray emission from W Comae if these models provide
an adequate description of the source.


\acknowledgments

We are grateful to the staff at the National Solar Thermal Test Facility,
who continue to support our science with enthusiasm and professionalism.
Thanks to Markus Boettcher and Anita Reimer for providing gamma-ray spectra
predicted by their leptonic and hadronic emission models of W Comae.
Thanks also to Teresa Spreitzer, Audry Alabiso, Joshua Boehm,
Nathan Kundtz, Dan Schuette, and Claude Th\'eoret.  This work was supported
in part by the National Science Foundation,
NSERC (the Natural Science and Engineering Research Council of Canada),
FQRNT (Fonds Qu\'eb\'ecois de la Recherche sur la Nature et les Technologies),
the Research Corporation, and the California Space Institute.


%
%


\begin{thebibliography}{199}


\bibitem[Aharonian(2000)]{aharonian00}
Aharonian, F. A.
   2000, New Astronomy, 5, 377

\bibitem[Aharonian(2001)]{aharonian01}
Aharonian, F. A.
   2001, in Proc. 27th International Cosmic Ray Conference
   (Hamburg, Germany)

\bibitem[Biermann(1997)]{biermann97}
Biermann, P. L.
   1997, J. Phys. G, 23, 1

\bibitem[Bloom \& Marscher(1996)]{bm96}
Bloom, S. D., \& Marscher, A. P.
   1996, \apj, 461, 657

\bibitem[Boettcher, Mukherjee \& Reimer(2002)]{bmr02}
Boettcher, M., Mukherjee, R., \& Reimer, A.
   2002, \apj, 581, 143

\bibitem[Boone \etal(2002)]{boonemrk421}
Boone, L. M. \etal
   2002, \apj, 579, L5

\bibitem[Cawley(1993)]{cawley93}
Cawley, M.
   1993, in \emph{Towards a Major Atmospheric Cherenkov Detector --- II},
   Calgary (ed. R. C. Lamb), 176

\bibitem[Coppi \& Aharonian(1999)]{ca99}
Coppi, P. S. \& Aharonian, F. A.
   1999, \apj, 521, L33

\bibitem[de Naurois \etal(2002)]{denaurois02}
de Naurois, M. \etal
   2002, ApJ, 566, 343

\bibitem[Dermer, Schlickeiser, \& Mastichiadis(1992)]{dsm92}
Dermer, C. D., Schlickeiser, R., \& Mastichiadis, A.
   1992, \aap, 256, L27

\bibitem[Dubovsky, Tinyakov \& Tkachev(2000)]{dtt00}
Dubovsky, S. L., Tinyakov, P. G., \& Tkachev, I. I.
   2000, \prl, 85, 1154

\bibitem[Hanna \& Mukherjee(2001)]{stlasernim}
Hanna, D. S. \& Mukherjee, R.
   2001, \NIMA, 482, 271

\bibitem[Hanna \etal(2002)]{stacee32nim}
Hanna, D. S. \etal
   2002, \NIMA, 491, 126

\bibitem[Hartman \etal(1999)]{hartman99}
Hartman, R. C., \etal
   1999, \apjs, 123, 79

\bibitem[Heck \etal(1999)]{corsika}
Heck, D., Knapp, J., Capdevielle, J. N., Schatz, G., \& Thou, T.
   1999, Rep. FZKA 6019, Forschungszentrum Karlsruhe

\bibitem[Horan \etal(2003)]{horan03}
Horan, D. \etal
   2003, in Proc. 28th International Cosmic Ray Conference
   (Tsukuba, Japan), OG 2.3 2567


\bibitem[Kalmykov \etal(1997)]{qgsjet}
Kalmykov, N. N. \etal
   1997, Nucl. Phys. B (Proc. Suppl.) 52B, 113

\bibitem[Konopelko \etal(2003)]{konopelko03}
Konopelko, A. \etal
   2003, in Proc. 28th International Cosmic Ray Conference
   (Tsukuba, Japan), OG 2.3 2611

\bibitem[Li \& Ma(1983)]{lima}
Li, T. P. \& Ma, Y. Q.
   1983, \apj, 272, 317

\bibitem[Malkan \& Stecker(2001)]{ms01}
Malkan, M. \& Stecker, F.
   2001, \apj, 555, 641


\bibitem[Mannheim(1993)]{mannheim93}
Mannheim, K.
   1993, \aap, 269, 67

\bibitem[Mannheim(1995)]{mannheim95}
Mannheim, K.
   1995, \app, 3, 295

\bibitem[Martin \& Ragan(2000)]{maddog}
Martin, J.-P. \& Ragan, K.
   2000, Proc. IEEE Nuclear Science Symposium, 12, 141.

\bibitem[M\"{u}cke \& Protheroe(2000)]{mp00}
M\"{u}cke, A., \& Protheroe, R. J.
   2000, in \emph{Towards a Major Atmospheric Cherenkov Detector --- VI},
   Salt Lake City (eds. B.L. Dingus, M. H. Salamon, D. B. Kieda),
   AIP Conference Proceedings vol. 515, p. 149

\bibitem[M\"{u}cke \etal(2003)]{mucke03}
M\"{u}cke, A., \etal
   2003, Astropart. Phys. 18, 593

\bibitem[Ning, Winston, \& O'Gallagher(1987)]{nwog87}
Ning, X., Winston, R., \& O'Gallagher, J.
   1987, Appl. Opt. 26, 300

\bibitem[Oser \etal(2001)]{osercrab}
Oser, S., \etal
   2001, \apj, 547, 949

\bibitem[Primack \etal(1999)]{primack99}
Primack, J. R., Bullock, J. S., Somerville, R. S., \& MacMinn, D.
   1999, Astropart. Phys. 11, 93

\bibitem[Primack \etal(2001)]{primack01}
Primack, J. R., \etal
   2001, in High-Energy Gamma-Ray Astronomy
      (eds. F. Aharonian \& H. J. V\"olk), AIP Conf. Series, 558, 463

\bibitem[Scalzo(2004)]{mythesis}
Scalzo, R. A.
   2004, Ph.D. thesis, University of Chicago

\bibitem[Scalzo \etal(2003)]{scalzo03}
Scalzo, R. A. \etal
   2003, in Proc. 28th International Cosmic Ray Conference
   (Tsukuba, Japan), OG 2.5 2799

\bibitem[Sikora, Begelman, \& Rees(1994)]{sbr94}
Sikora, M., Begelman, M. C., \& Rees, M. J.
   1994, \apj, 421, 153

\bibitem[Tagliaferri \etal(2000)]{tagliaferri00}
Tagliaferri, G. \etal
   2000, \aap, 354, 431

\bibitem[Th\'eoret(2001)]{theoret-thesis}
Th\'eoret, C. G.
   2001, Ph.D. thesis, McGill University

\bibitem[Tripathi(2002)]{tripathi02}
Tripathi, S. M. \etal
   2002, BAAS, 34, 676

\bibitem[Urry \& Padovani(1995)]{up95}
Urry, C. M. \& Padovani, P.
   1995, \pasp, 107, 715

\bibitem[Weekes(1988)]{weekes88}
Weekes, T. C.
   1988, \physrep, 160, 1

\bibitem[Weekes \etal(1989)]{weekes89}
Weekes, T. C. \etal
   1989, \apj, 342, 379

\bibitem[Zweerink \etal(2003)]{zweerink03}
Zweerink, J. A. \etal
   2003, in Proc. 28th International Cosmic Ray Conference
   (Tsukuba, Japan), OG 2.5 2795



\end{thebibliography}
\end{document}